\documentclass[conference]{IEEEtran}

\IEEEoverridecommandlockouts

\usepackage{multirow}
\usepackage[table,xcdraw]{xcolor}
\usepackage{threeparttable}

\usepackage{changes}
\definechangesauthor[name={Wei}, color=blue]{YN}
\definechangesauthor[name={Wei}, color=red]{DEL}

\usepackage{cite}
\usepackage{amsmath,amssymb,amsfonts}
\usepackage{algorithmic}
\usepackage{graphicx}
\usepackage{textcomp}
\usepackage[table,xcdraw]{xcolor}
\usepackage{soul}
\usepackage{threeparttable}

\definecolor{amethyst}{rgb}{0.6, 0.4, 0.8}

\def\BibTeX{{\rm B\kern-.05em{\sc i\kern-.025em b}\kern-.08em
    T\kern-.1667em\lower.7ex\hbox{E}\kern-.125emX}}
\begin{document}

\title{Unlocking Mental Health: Exploring College Students' Well-being through Smartphone Behaviors
}

\author{\IEEEauthorblockN{1\textsuperscript{st} Wei Xuan}
\IEEEauthorblockA{\textit{Department of Economics} \\
\textit{University of Southern California}\\
Los Angeles, CA, USA \\
wxuan@usc.edu}
\and
\IEEEauthorblockN{2\textsuperscript{nd} Meghna Roy Chowdhury, 3\textsuperscript{rd} Yi Ding}
\IEEEauthorblockA{\textit{Electrical and Computer Engineering} \\
\textit{Purdue University}\\
West Lafayette, IN, USA \\
mroycho@purdue.edu, yiding@purdue.edu}
\and
\IEEEauthorblockN{4\textsuperscript{th} Yixue Zhao}
\IEEEauthorblockA{\textit{Information Sciences Institute} \\
\textit{University of Southern California}\\
Arlington, VA, USA \\
yzhao@isi.edu}
}

\maketitle

\begin{abstract}
The global mental health crisis is a pressing concern, with college students particularly vulnerable to rising mental health disorders. The widespread use of smartphones among young adults, while offering numerous benefits, has also been linked to negative outcomes such as addiction and regret, significantly impacting well-being. Leveraging the longest longitudinal dataset collected over four college years through passive mobile sensing, this study is the first to examine the relationship between students' smartphone unlocking behaviors and their mental health at scale in real-world settings. We provide the first evidence demonstrating the predictability of phone unlocking behaviors for mental health outcomes based on a large dataset, highlighting the potential of these novel features for future predictive models. Our findings reveal important variations in smartphone usage across genders and locations, offering a deeper understanding of the interplay between digital behaviors and mental health. We highlight future research directions aimed at mitigating adverse effects and promoting digital well-being in this population.
\end{abstract}


\vspace{-2pt}
\section{Introduction}\label{intro}
\vspace{-2pt}
College years are pivotal for young adults' personal, social, and academic development, often accompanied by significant stressors \cite{magolda2006intellectual,anderson2012college}. Mobile devices have become integral to college students' daily lives, with U.S. students averaging 8–10 hours of smartphone use per day \cite{roberts2014invisible}. 
A considerable body of literature suggests that excessive smartphone usage interferes with daily activities and diminishes mental health \cite{ames2013managing,cho2021reflect,ruensuk2022sad,lin2016association}. However, most prior work overlooks the multifaceted nature of smartphone use. When used mindfully, smartphones can be powerful tools that enhance productivity and well-being \cite{buchi2024digital, zhao2024digital}. Furthermore, smartphones' impact likely varies across user groups and contexts, highlighting the need for a more nuanced understanding of its relationship with mental health.

To gain a comprehensive understanding of college students' smartphone usage and its association with mental well-being, we utilized data from the {College Experience Study} (CES) dataset~\cite{nepal2024capturing}, the longest-running mobile sensing study to date. This study provides a rich dataset of passive mobile sensing data and Ecological Momentary Assessment (EMA) surveys collected from over 200 Dartmouth students between 2017 and 2022. 
Our analysis specifically focuses on the fundamental smartphone behavior of \textit{unlocking}, considering both its \emph{frequency} and \emph{duration}, to explore its relationship with students' mental health as measured by the Patient Health Questionnaire-4 (PHQ4)\cite{lowe20104,kroenke2009ultra}.
We focus on \emph{unlocking} behaviors due to empirical evidence linking them to mental health and their coarse-grained nature, which allows prediction without compromising user privacy (e.g., revealing app usage). 

This is the first large-scale study based on over 210,000 data points collected throughout entire college years that utilizes {real-world} unlocking data to explore its predictability for mental health, moving beyond prior work that has relied on indirect methods to study phone usage~\cite{abdrabou2024eyegaze, harbach2014survey, li2019swipevlock}. 
We first conduct correlation analysis to assess the feasibility of using smartphone unlocking behaviors to predict students' mental health status. To guide the granularity of future predictive models, we further explored the impact of gender and location differences (e.g., study spaces, social venues, and homes) on the correlations.
Following the standardization approach for PHQ4 Score in \cite{wicke2022update}, we categorize the PHQ4 Score into four groups---“Normal,” “Mild,” “Moderate,” and “Severe”---to facilitate easier interpretation. Due to the discrete nature of these outcome variables, we employ multinomial logistic regression~\cite{bohning1992multinomial} to analyze the relationships between smartphone unlocking behaviors and mental well-being. Rather than using a single global model for the entire population, we apply separate multinomial logistic regression models for different genders and locations to capture the dataset's heterogeneity.

Our analysis uncovers nuanced phenomena in terms of gender differences: for male students, increased phone usage is linked to poorer mental health, while for female students, it appears to have a positive impact, suggesting fundamentally different dynamics in how phone usage affects well-being across genders. 
These findings provide insights and actionable items for future work to build predictive models and design digital well-being interventions, as discussed in Section~\ref{result}. 

This paper makes these contributions: 
1) We conduct the first large-scale study of real-world smartphone unlocking behaviors and mental health among college students, encompassing both iOS and Android users; 
2) We provide the first empirical evidence on the predictability of novel smartphone unlocking features, opening up future directions in predicting mental health; 
3) We demonstrate the gender differences and location variations in students' smartphone behaviors and their relationships with mental health, providing actionable insights for building predictive models;
4) We open-source the data analysis pipeline and artifacts to foster future research~\cite{OurRepo}.


\section{Empirical Study Design}\label{design}

\subsection{Research Questions}

To investigate the nuanced relationship between college students' smartphone behaviors and mental health, as well as to guide future predictive models on mental health outcomes, we address the following key research questions (RQs).
\begin{itemize}
	\item \textbf{RQ$_1$} -- Do college students of different genders interact with their smartphones differently across college years?
	\item \textbf{RQ$_2$} -- Are there any correlations between students' smartphone behaviors and mental health? Do gender or location differences change these correlations? 
	\item \textbf{RQ$_3$} -- Can we build predictive models of students' mental health with smartphone behaviors? Do predictive patterns differ across genders and locations?
\end{itemize}

\subsection{Dataset Overview and Pre-processing}\label{data}
\vspace{-2pt}

To answer the RQs, we utilize the CES dataset \cite{nepal2024capturing}, the most extensive longitudinal mobile sensing study to date. This dataset spans from 2017 to 2022 and includes 215 first-year undergraduate students from Dartmouth College in the United States, with 146 (67.9\%) female and 69 (32.1\%) male participants.
From the dataset, we select the most relevant features to represent smartphone behaviors, namely, the number of phone unlocks (\textit{Unlock Number}) and the duration of phone unlocks (\textit{Unlock Duration}) on a given day.
Students' mental health was assessed using the \textit{PHQ4 Score} from the EMA surveys based on the PHQ9 and the Generalized Anxiety Disorder-7 (GAD7)~\cite{nepal2024capturing,williams2014gad}. PHQ4 Score ranges from 0 to 12, where higher scores denote greater levels of depression and anxiety.
The EMA surveys are delivered randomly once per week through a mobile app. In total, the dataset contains 213,408 data entries collected for unlocking behaviors and 34,581 data entries for PHQ4 Score across all participants.


We augment the dataset by first introducing a new feature to represent the average time spent on each unlock on a given day (\textit{Duration per Unlock}). To align the data with the PHQ4 Score, which reflects mental health status over the past two weeks, we introduce three additional features and calculate their values: the moving averages of {Unlock Duration}, {Unlock Number}, and {Duration per Unlock} over the same two-week period. 

To avoid tainted results, we exclude outlier data points in our analyses 
based on the following criteria: (1) data points where the {Unlock Duration} exceeds 16 hours; (2) students whose maximum {Unlock Number} or {Unlock Duration} is 0; and (3) students whose data shows no variation, defined as a standard deviation of 0 for {Unlock Number}, {Unlock Duration}, or {PHQ4 Score} throughout the study period.
This process resulted in 214 students with 213,360 data points in total.


\subsection{Unlocking Behavior Patterns}
\label{sec:data:distribution}
\vspace{-2pt}
To first understand unlocking behavior patterns among college students (\textbf{RQ$_1$}), we calculate the detailed statistics (e.g., mean, standard deviation) of {Unlock Number}, {Unlock Duration}, and {Duration per Unlock} for each student. We then plot the distributions of these features across all students and further stratify the distributions by gender to reveal potential differences between male and female students.


\subsection{Pearson Correlation Analysis}
\label{sec:data:correlation}
\vspace{-2pt}
To understand whether unlocking behaviors are promising features to predict mental health, we investigate the direction and strength of correlations between unlocking behaviors and students' mental health (\textbf{RQ$_2$}) by calculating the Pearson correlation coefficients between PHQ4 Scores and {Unlock Duration}, {Unlock Number}, and {Duration per Unlock}. These coefficients are stratified by gender and location to uncover potential variations, providing further insights.

\subsection{Multinomial Logistic Regression Model}
\label{sec:data:regression}
\vspace{-2pt}
To explore the extent to which unlock features can predict students' mental health (\textbf{RQ$_3$}), we formulate it as a classification problem by categorizing the outcome variable {PHQ4 Score} into four groups using standardization approach~\cite{wicke2022update}: (1) ``Normal'' (0–2); (2) ``Mild'' (3-5); (3) ``Moderate'' (6-8); (4) ``Severe'' (9-12). This approach accommodates the subjectivity inherent in PHQ4 Scores, as students may rate the same mental status with different scores. By allowing variation within each group, we reduce the sensitivity and improve the robustness of the classification. Given the discrete and multi-class nature of the outcome variables, we use multinomial logistic regression~\cite{bohning1992multinomial} to model the relationship between unlocking behaviors and the likelihood of belonging to different mental status groups. This method is chosen for its suitability in handling categorical outcomes and its inherent linearity, ensuring interpretability and computational efficiency.

\vspace{-2pt}
\section{Results and Lessons Learned}
\label{result}
\vspace{-2pt}
\subsection{Insights for Unlocking Behavior}
\vspace{-2pt}
\begin{figure}[b]
\centering
\includegraphics[width=\linewidth]{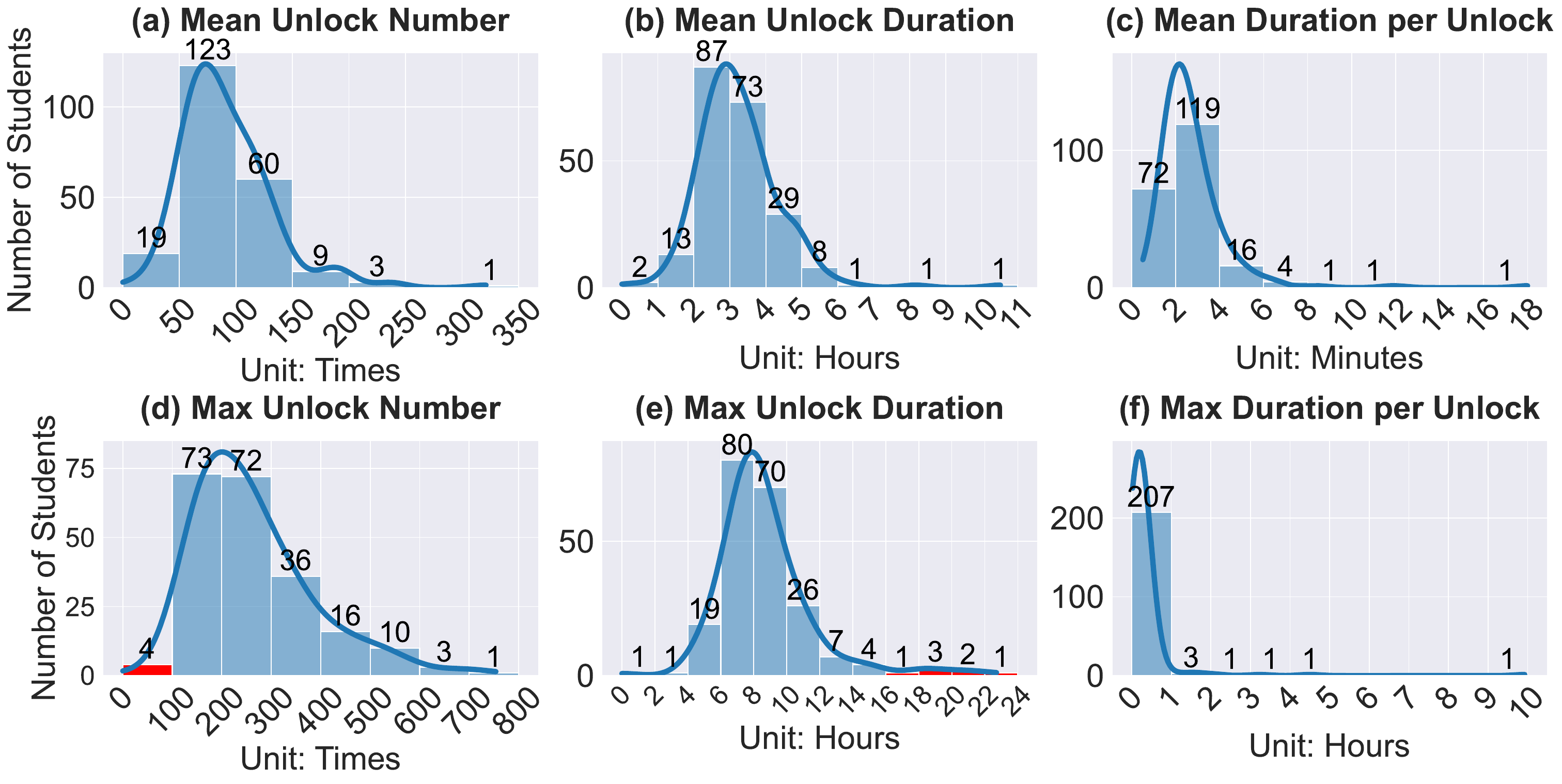}
\caption{Distributions of Unlock Number, Unlock Duration, and Duration per Unlock at the individual level with mean values (top) and maximum values (bottom). Excluded data points described in Section \ref{data} are shown in red.}
\label{fig1}
\end{figure}

\begin{figure}[t]
\centering
\includegraphics[width=\linewidth]{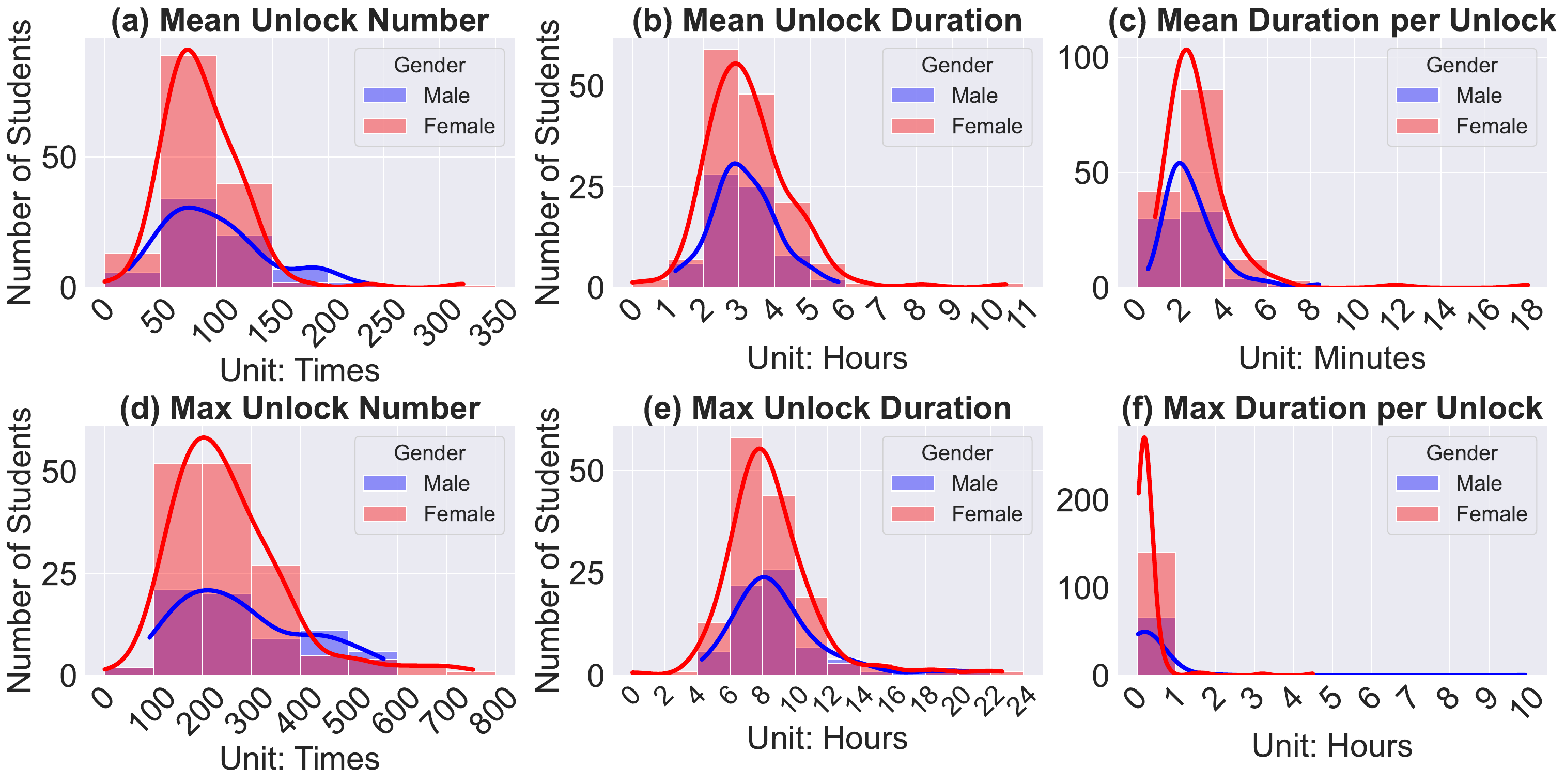}
\caption{Gender differences highlighted based on the same data shown in Fig.~\ref{fig1}}
\label{fig2}
\end{figure}


To answer \textbf{RQ$_1$}, we analyze the data distribution as discussed in Section~\ref{sec:data:distribution}. Fig.~\ref{fig1} highlights the mean and maximum statistics of the results. For instance, Fig.\ref{fig1}-a (Mean Unlock Number) shows the distribution of the mean value of each student's Unlock Number during the study period.

\textbf{Overall Pattern}. On average, most students unlock their phones 50–100 times daily (Fig.\ref{fig1}-a) totaling 2–4 hours (Fig.\ref{fig1}-b), and each unlock session averages 2-4 minutes (Fig.\ref{fig1}-c). On high-usage days, most students unlock their phones 100-300 times (Fig.\ref{fig1}-d)
totaling 6–10 hours. Interestingly, for most students, the duration per unlock typically falls under 4 minutes (Fig.\ref{fig1}-c), and the majority of students (96\%) use their phones under 1 hour per unlock even on the highest-usage days (Fig.\ref{fig1}-f). This suggests that students primarily engage with their phones for brief, fleeting interactions rather than extended, focused tasks, likely to satisfy momentary curiosity or to simply mindlessly unlock rather than sustained activities.

\textbf{Gender Difference}. As shown in Fig. \ref{fig2}, notable gender differences in phone usage emerge. Male students display a broader range of unlock frequencies, indicating more diverse usage habits. In contrast, female students exhibit more consistent patterns but spend significantly more time per unlock. This behavior suggests that female students may engage more deeply with their phones during each interaction, potentially reflecting a more intentional and focused usage style.

\vspace{-2pt}
\subsection{Unlocking Behavior and Mental Well-being Correlation}
\vspace{-2pt}


\begin{table}[b]
\centering
\caption{Pearson Correlation with PHQ4 Score across Genders}
\begin{tabular}{|l|c|c|c|}
\hline
\textbf{}                    & \textbf{Overall}                                                                                  & \textbf{Male}                                                                                    & \textbf{Female}                                                                                   \\ \hline
\textbf{Unlock Duration}     & {\color[HTML]{3166FF} \begin{tabular}[c]{@{}c@{}}0.0321 \\      (p\textless{}0.01)\end{tabular}}  & {\color[HTML]{3166FF} \begin{tabular}[c]{@{}c@{}}0.2131 \\      (p\textless{}0.01)\end{tabular}} & {\color[HTML]{FE0000} \begin{tabular}[c]{@{}c@{}}-0.0388 \\      (p\textless{}0.01)\end{tabular}} \\ \hline
\textbf{Unlock Number}       & {\color[HTML]{FE0000} \begin{tabular}[c]{@{}c@{}}-0.0107 \\      (p\textless{}0.01)\end{tabular}} & {\color[HTML]{3166FF} \begin{tabular}[c]{@{}c@{}}0.0964 \\      (p\textless{}0.01)\end{tabular}} & {\color[HTML]{FE0000} \begin{tabular}[c]{@{}c@{}}-0.0743 \\      (p\textless{}0.01)\end{tabular}} \\ \hline
\textbf{Duration per Unlock} & {\color[HTML]{3166FF} \begin{tabular}[c]{@{}c@{}}0.0380 \\      (p\textless{}0.01)\end{tabular}}  & {\color[HTML]{3166FF} \begin{tabular}[c]{@{}c@{}}0.0778 \\      (p\textless{}0.01)\end{tabular}} & {\color[HTML]{3166FF} \begin{tabular}[c]{@{}c@{}}0.0284 \\      (p\textless{}0.01)\end{tabular}}  \\ \hline
\end{tabular}
\label{corre}
\begin{tablenotes}
\item Positive coefficients are in blue, while negative coefficients are in red. 
\end{tablenotes}
\end{table}

\begin{table}[t]
\centering
\caption{Pearson Correlation with PHQ4 Score at Different Locations}
\begin{tabular}{|l|l|l|l|l|}
\hline
                                     &                          & \textbf{Overall}                                                                               & \textbf{Male}                                                                                  & \textbf{Female}                                                                               \\ \hline
                                     & \textbf{Duration/Unlock} & {\color[HTML]{3166FF} \begin{tabular}[c]{@{}l@{}}0.0115\\ (p\textgreater{}0.1)\end{tabular}}   & {\color[HTML]{3166FF} \begin{tabular}[c]{@{}l@{}}0.0791\\ (p\textless{}0.01)\end{tabular}}     & {\color[HTML]{3166FF} \begin{tabular}[c]{@{}l@{}}0.0043\\ (p\textgreater{}0.1)\end{tabular}}  \\ \cline{2-5} 
                                     & \textbf{Duration}        & {\color[HTML]{FE0000} \begin{tabular}[c]{@{}l@{}}-0.0125\\ (p\textless{}0.05)\end{tabular}}    & {\color[HTML]{3166FF} \begin{tabular}[c]{@{}l@{}}0.0296\\ (p\textless{}0.01)\end{tabular}}     & {\color[HTML]{FE0000} \begin{tabular}[c]{@{}l@{}}-0.0265\\ (p\textless{}0.01)\end{tabular}}   \\ \cline{2-5} 
\multirow{-3}{*}{\textbf{Food}}      & \textbf{Number}          & {\color[HTML]{FE0000} \begin{tabular}[c]{@{}l@{}}-0.0197\\ (p\textless{}0.01)\end{tabular}}    & {\color[HTML]{3166FF} \begin{tabular}[c]{@{}l@{}}0.0214\\ (p\textless{}0.05)\end{tabular}}     & {\color[HTML]{FE0000} \begin{tabular}[c]{@{}l@{}}-0.0408\\ (p\textless{}0.01)\end{tabular}}   \\ \hline
                                     & \textbf{Duration/Unlock} & {\color[HTML]{3166FF} \begin{tabular}[c]{@{}l@{}}0.038\\ (p\textless{}0.01)\end{tabular}}      & {\color[HTML]{3166FF} \begin{tabular}[c]{@{}l@{}}0.0479\\ (p\textless{}0.01)\end{tabular}}     & {\color[HTML]{3166FF} \begin{tabular}[c]{@{}l@{}}0.0343\\ (p\textless{}0.01)\end{tabular}}    \\ \cline{2-5} 
                                     & \textbf{Duration}        & {\color[HTML]{3166FF} \begin{tabular}[c]{@{}l@{}}0.0014\\ (p\textgreater{}0.1)\end{tabular}}   & {\color[HTML]{FE0000} \begin{tabular}[c]{@{}l@{}}-0.0055\\ (p\textgreater{}0.1)\end{tabular}}  & {\color[HTML]{3166FF} \begin{tabular}[c]{@{}l@{}}0.009\\ (p\textgreater{}0.1)\end{tabular}}   \\ \cline{2-5} 
\multirow{-3}{*}{\textbf{Study}}     & \textbf{Number}          & {\color[HTML]{FE0000} \begin{tabular}[c]{@{}l@{}}-0.0151\\ (p\textless{}0.01)\end{tabular}}    & {\color[HTML]{FE0000} \begin{tabular}[c]{@{}l@{}}-0.0218\\ (p\textless{}0.05)\end{tabular}}    & {\color[HTML]{FE0000} \begin{tabular}[c]{@{}l@{}}-0.0077\\ (p\textgreater{}0.1)\end{tabular}} \\ \hline
                                     & \textbf{Duration/Unlock} & {\color[HTML]{FE0000} \begin{tabular}[c]{@{}l@{}}-0.0005\\ (p\textgreater{}0.1)\end{tabular}}  & {\color[HTML]{3166FF} \begin{tabular}[c]{@{}l@{}}0.0371\\ (p\textless{}0.05)\end{tabular}}     & {\color[HTML]{FE0000} \begin{tabular}[c]{@{}l@{}}-0.0082\\ (p\textgreater{}0.1)\end{tabular}} \\ \cline{2-5} 
                                     & \textbf{Duration}        & {\color[HTML]{FE0000} \begin{tabular}[c]{@{}l@{}}-0.01\\ (p\textless{}0.1)\end{tabular}}       & {\color[HTML]{FE0000} \begin{tabular}[c]{@{}l@{}}-0.0046\\ (p\textgreater{}0.1)\end{tabular}}  & {\color[HTML]{FE0000} \begin{tabular}[c]{@{}l@{}}-0.0073\\ (p\textgreater{}0.1)\end{tabular}} \\ \cline{2-5} 
\multirow{-3}{*}{\textbf{Social}}    & \textbf{Number}          & {\color[HTML]{FE0000} \begin{tabular}[c]{@{}l@{}}-0.0157\\ (p\textless{}0.01)\end{tabular}}    & {\color[HTML]{FE0000} \begin{tabular}[c]{@{}l@{}}-0.0057\\ (p\textgreater{}0.1)\end{tabular}}  & {\color[HTML]{FE0000} \begin{tabular}[c]{@{}l@{}}-0.0181\\ (p\textless{}0.01)\end{tabular}}   \\ \hline
                                     & \textbf{Duration/Unlock} & {\color[HTML]{FE0000} \begin{tabular}[c]{@{}l@{}}0.02478\\ (p\textless{}0.01)\end{tabular}}    & {\color[HTML]{FE0000} \begin{tabular}[c]{@{}l@{}}-0.00979\\ (p\textgreater{}0.1)\end{tabular}} & {\color[HTML]{3166FF} \begin{tabular}[c]{@{}l@{}}0.04242\\ (p\textless{}0.01)\end{tabular}}   \\ \cline{2-5} 
                                     & \textbf{Duration}        & {\color[HTML]{FE0000} \begin{tabular}[c]{@{}l@{}}-0.00215\\ (p\textgreater{}0.1)\end{tabular}} & {\color[HTML]{FE0000} \begin{tabular}[c]{@{}l@{}}-0.01371\\ (p\textgreater{}0.1)\end{tabular}} & {\color[HTML]{3166FF} \begin{tabular}[c]{@{}l@{}}0.00987\\ (p\textgreater{}0.1)\end{tabular}} \\ \cline{2-5} 
\multirow{-3}{*}{\textbf{Dormitory}} & \textbf{Number}          & {\color[HTML]{FE0000} \begin{tabular}[c]{@{}l@{}}-0.01505\\ (p\textless{}0.01)\end{tabular}}   & {\color[HTML]{FE0000} \begin{tabular}[c]{@{}l@{}}-0.00768\\ (p\textgreater{}0.1)\end{tabular}} & {\color[HTML]{FE0000} \begin{tabular}[c]{@{}l@{}}-0.01684\\ (p\textless{}0.01)\end{tabular}}  \\ \hline
                                     & \textbf{Duration/Unlock} & {\color[HTML]{3166FF} \begin{tabular}[c]{@{}l@{}}0.0181\\ (p\textless{}0.01)\end{tabular}}     & {\color[HTML]{3166FF} \begin{tabular}[c]{@{}l@{}}0.0562\\ (p\textless{}0.01)\end{tabular}}     & {\color[HTML]{3166FF} \begin{tabular}[c]{@{}l@{}}0.0145\\ (p\textless{}0.05)\end{tabular}}    \\ \cline{2-5} 
                                     & \textbf{Duration}        & {\color[HTML]{3166FF} \begin{tabular}[c]{@{}l@{}}0.0188\\ (p\textless{}0.01)\end{tabular}}     & {\color[HTML]{3166FF} \begin{tabular}[c]{@{}l@{}}0.1474\\ (p\textless{}0.01)\end{tabular}}     & {\color[HTML]{FE0000} \begin{tabular}[c]{@{}l@{}}-0.0334\\ (p\textless{}0.01)\end{tabular}}   \\ \cline{2-5} 
\multirow{-3}{*}{\textbf{Home}}      & \textbf{Number}          & {\color[HTML]{3166FF} \begin{tabular}[c]{@{}l@{}}0.0033\\ (p\textgreater{}0.1)\end{tabular}}   & {\color[HTML]{3166FF} \begin{tabular}[c]{@{}l@{}}0.092\\ (p\textless{}0.01)\end{tabular}}      & {\color[HTML]{FE0000} \begin{tabular}[c]{@{}l@{}}-0.047\\ (p\textless{}0.01)\end{tabular}}    \\ \hline
\end{tabular}
\label{corre_loc}
\begin{tablenotes}
\item Positive coefficients are in blue, while negative coefficients are in red. 
\end{tablenotes}
\vspace{-2pt}
\end{table}

\textbf{Overall Correlations}. To answer \textbf{RQ$_2$}, we conduct correlation analysis as described in Section~\ref{sec:data:correlation} and the results are shown in Table~\ref{corre}.
Notably, all p-values for the correlation coefficients are well below 0.01, indicating associations exist between phone usage patterns and mental health. This finding highlights the potential of our novel unlocking features as effective predictors of mental health outcomes. Column 1 reveals an intriguing pattern: longer phone usage durations correlate with poorer mental health outcomes, whereas higher unlock frequencies are associated with better mental health. These findings underscore the importance of considering both frequency and duration when assessing the impact of smartphone use on mental health, and call for further studies to identify the root causes of these interesting phenomena.

\textbf{Gender Difference}. 
Table \ref{corre} reveals significant gender differences in phone usage. For males, both unlock duration and frequency are positively correlated with PHQ4 Scores, indicating a link to poorer mental health. In contrast, females exhibit the opposite trend, suggesting higher phone usage may be associated with better well-being or act as a coping mechanism. Additionally, the magnitude of correlation coefficients is notably larger for male students, suggesting they are more sensitive to the mental health impacts of phone usage. These findings imply that male students may engage with their phones in less constructive ways than their female peers, with higher interaction levels correlating more strongly with adverse mental health outcomes, highlighting the need for personalized predictive models in different user groups.

\textbf{Location Difference}. Table \ref{corre_loc} shows the correlations between unlocking behaviors and mental well-being at different locations, revealing fine-grained insights to guide further studies. For example, unlocking frequency and duration are both negatively correlated with PHQ4 Score (better mental health) at food and social places, suggesting beneficial effects in social contexts. In contrast, these behaviors correlate with poorer mental health at home, possibly reflecting less meaningful usage when in private. These findings emphasize the importance of context when examining phone usage and mental health.

\subsection{Insights from Multinomial Logistic Regression}
\vspace{-2pt}

\begin{table}[t]
\centering
\caption{Coefficients Results of Multinomial Logistic Regression}
\label{logit}
\begin{tabular}{|l|l|l|l|l|}
\hline
                               & \textbf{Features} & \textbf{Mild}                                                                                          & \textbf{Moderate}                                                                                       & \textbf{Severe}                                                                                        \\ \hline
                               & \textbf{Duration} & {\color[HTML]{3166FF} \begin{tabular}[c]{@{}l@{}}0.1169\\  (p\textless{}0.01)\end{tabular}}    & {\color[HTML]{FE0000} \begin{tabular}[c]{@{}l@{}}-0.0339\\  (p\textless{}0.01)\end{tabular}}    & {\color[HTML]{3166FF} \begin{tabular}[c]{@{}l@{}}0.1575\\  (p\textless{}0.01)\end{tabular}}    \\ \cline{2-5} 
\multirow{-2}{*}{\textbf{Overall}}      & \textbf{Number}   & {\color[HTML]{3166FF} \begin{tabular}[c]{@{}l@{}}0.0903\\  (p\textless{}0.01)\end{tabular}}    & {\color[HTML]{FE0000} \begin{tabular}[c]{@{}l@{}}-0.0378\\  (p\textless{}0.01)\end{tabular}}    & {\color[HTML]{FE0000} \begin{tabular}[c]{@{}l@{}}-0.1270\\  (p\textless{}0.01)\end{tabular}}   \\ \hline
                               & \textbf{Duration} & {\color[HTML]{3166FF} \begin{tabular}[c]{@{}l@{}}0.2398\\  (p\textless{}0.01)\end{tabular}}    & {\color[HTML]{3166FF} \begin{tabular}[c]{@{}l@{}}0.0298\\  (p\textgreater{}0.1)\end{tabular}}   & {\color[HTML]{3166FF} \begin{tabular}[c]{@{}l@{}}0.5741\\  (p\textless{}0.01)\end{tabular}}    \\ \cline{2-5} 
\multirow{-2}{*}{\textbf{Male}}         & \textbf{Number}   & {\color[HTML]{3166FF} \begin{tabular}[c]{@{}l@{}}0.2215\\  (p\textless{}0.01)\end{tabular}}    & {\color[HTML]{FE0000} \begin{tabular}[c]{@{}l@{}}-0.1896\\  (p\textless{}0.01)\end{tabular}}    & {\color[HTML]{FE0000} \begin{tabular}[c]{@{}l@{}}-0.1731\\  (p\textless{}0.01)\end{tabular}}   \\ \hline
                               & \textbf{Duration} & {\color[HTML]{3166FF} \begin{tabular}[c]{@{}l@{}}0.1000\\  (p\textless{}0.01)\end{tabular}}    & {\color[HTML]{FE0000} \begin{tabular}[c]{@{}l@{}}-0.0386\\  (p\textless{}0.01)\end{tabular}}    & {\color[HTML]{FE0000} \begin{tabular}[c]{@{}l@{}}-0.0053\\  (p\textgreater{}0.1)\end{tabular}} \\ \cline{2-5} 
\multirow{-2}{*}{\textbf{Female}}       & \textbf{Number}   & {\color[HTML]{FE0000} \begin{tabular}[c]{@{}l@{}}-0.0810\\  (p\textless{}0.05)\end{tabular}}   & {\color[HTML]{3166FF} \begin{tabular}[c]{@{}l@{}}0.0147\\  (p\textgreater{}0.1)\end{tabular}}   & {\color[HTML]{FE0000} \begin{tabular}[c]{@{}l@{}}-0.2997\\  (p\textless{}0.01)\end{tabular}}   \\ \hline
                               & \textbf{Duration} & {\color[HTML]{3166FF} \begin{tabular}[c]{@{}l@{}}0.0547\\  (p\textgreater{}0.1)\end{tabular}}  & {\color[HTML]{FE0000} \begin{tabular}[c]{@{}l@{}}-0.0283\\  (p\textgreater{}0.1)\end{tabular}}  & {\color[HTML]{3166FF} \begin{tabular}[c]{@{}l@{}}0.1400\\  (p\textless{}0.01)\end{tabular}}    \\ \cline{2-5} 
\multirow{-2}{*}{\textbf{Food Places}}   & \textbf{Number}   & {\color[HTML]{FE0000} \begin{tabular}[c]{@{}l@{}}-0.0059\\  (p\textgreater{}0.1)\end{tabular}} & {\color[HTML]{FE0000} \begin{tabular}[c]{@{}l@{}}-0.0444\\  (p\textless{}0.05)\end{tabular}}    & {\color[HTML]{FE0000} \begin{tabular}[c]{@{}l@{}}-0.3357\\  (p\textless{}0.01)\end{tabular}}   \\ \hline
                               & \textbf{Duration} & {\color[HTML]{3166FF} \begin{tabular}[c]{@{}l@{}}0.0586\\  (p\textgreater{}0.1)\end{tabular}}  & {\color[HTML]{FE0000} \begin{tabular}[c]{@{}l@{}}-0.0294\\  (p\textgreater{}0.1)\end{tabular}}  & {\color[HTML]{3166FF} \begin{tabular}[c]{@{}l@{}}0.1417\\  (p\textless{}0.01)\end{tabular}}    \\ \cline{2-5} 
\multirow{-2}{*}{\textbf{Social Places}} & \textbf{Number}   & {\color[HTML]{FE0000} \begin{tabular}[c]{@{}l@{}}-0.0062\\  (p\textgreater{}0.1)\end{tabular}} & {\color[HTML]{FE0000} \begin{tabular}[c]{@{}l@{}}-0.0450\\  (p\textless{}0.05)\end{tabular}}    & {\color[HTML]{FE0000} \begin{tabular}[c]{@{}l@{}}-0.3365\\  (p\textless{}0.01)\end{tabular}}   \\ \hline
                               & \textbf{Duration} & {\color[HTML]{3166FF} \begin{tabular}[c]{@{}l@{}}0.0060\\  (p\textgreater{}0.1)\end{tabular}}  & {\color[HTML]{FE0000} \begin{tabular}[c]{@{}l@{}}-0.0688\\  (p\textless{}0.01)\end{tabular}}    & {\color[HTML]{3166FF} \begin{tabular}[c]{@{}l@{}}0.0864\\  (p\textgreater{}0.1)\end{tabular}}  \\ \cline{2-5} 
\multirow{-2}{*}{\textbf{Study Places}}        & \textbf{Number}   & {\color[HTML]{3166FF} \begin{tabular}[c]{@{}l@{}}0.0111\\  (p\textgreater{}0.1)\end{tabular}}  & {\color[HTML]{FE0000} \begin{tabular}[c]{@{}l@{}}-0.0306\\  (p\textless{}0.05)\end{tabular}} & {\color[HTML]{FE0000} \begin{tabular}[c]{@{}l@{}}-0.2858\\  (p\textless{}0.01)\end{tabular}}   \\ \hline
                               & \textbf{Duration} & {\color[HTML]{3166FF} \begin{tabular}[c]{@{}l@{}}0.1253\\  (p\textless{}0.01)\end{tabular}}    & {\color[HTML]{FE0000} \begin{tabular}[c]{@{}l@{}}-0.0583\\  (p\textless{}0.01)\end{tabular}}    & {\color[HTML]{3166FF} \begin{tabular}[c]{@{}l@{}}0.0735\\  (p\textless{}0.05)\end{tabular}} \\ \cline{2-5} 
\multirow{-2}{*}{\textbf{Home}}         & \textbf{Number}   & {\color[HTML]{3166FF} \begin{tabular}[c]{@{}l@{}}0.0845\\  (p\textless{}0.01)\end{tabular}}    & {\color[HTML]{FE0000} \begin{tabular}[c]{@{}l@{}}-0.0455\\  (p\textless{}0.01)\end{tabular}}    & {\color[HTML]{FE0000} \begin{tabular}[c]{@{}l@{}}-0.0965\\  (p\textless{}0.05)\end{tabular}}   \\ \hline
                               & \textbf{Duration} & {\color[HTML]{3166FF} \begin{tabular}[c]{@{}l@{}}0.0032\\  (p\textgreater{}0.1)\end{tabular}}  & {\color[HTML]{FE0000} \begin{tabular}[c]{@{}l@{}}-0.0789\\  (p\textless{}0.01)\end{tabular}}    & {\color[HTML]{FE0000} \begin{tabular}[c]{@{}l@{}}-0.0187\\  (p\textgreater{}0.1)\end{tabular}} \\ \cline{2-5} 
\multirow{-2}{*}{\textbf{Dormitories}}    & \textbf{Number}   & {\color[HTML]{3166FF} \begin{tabular}[c]{@{}l@{}}0.0687\\  (p\textless{}0.05)\end{tabular}}    & {\color[HTML]{FE0000} \begin{tabular}[c]{@{}l@{}}-0.0019\\  (p\textgreater{}0.1)\end{tabular}}  & {\color[HTML]{FE0000} \begin{tabular}[c]{@{}l@{}}-0.1317\\  (p\textless{}0.05)\end{tabular}}   \\ \hline
\end{tabular}
\begin{tablenotes}
\item Positive coefficients are in blue, while negative coefficients are in red. 
\end{tablenotes}
\vspace{-2pt}
\end{table}
\textbf{Impact of Phone Usage is Multifaceted}. Table \ref{logit} answers \textbf{RQ$_3$} as described in Section~\ref{sec:data:regression}
with ``Normal'' as the baseline group, uncovering interesting patterns. For instance, the overall results (first row) suggest that longer unlock durations are associated with higher likelihoods of being in the ``Mild'' or ``Severe'' mental health categories compared to the ``Normal'' group. However, for ``Moderate'' cases, longer unlock durations are more likely to be observed in individuals classified under the ``Normal'' mental status.
This finding suggests that the impact of phone usage on mental health is nuanced and not in one single direction. Phone usage is not inherently beneficial or harmful; its effect varies by individual. Wise phone usage may support better mental well-being, increasing the chances of being in less severe categories, while excessive or addictive use can worsen mental health outcomes.

\textbf{Gender Differences in Predictive Patterns}. 
Among males, longer unlock durations are associated with a higher likelihood of being in the ``Severe'' mental health category, whereas for females, longer durations are more likely to be linked to ``Normal'' state rather than ``Severe''. 
Notably, male students exhibit greater sensitivity to unlocking behaviors with a higher probability linking longer unlock duration to ``Severe'' mental status (coefficient\textgreater{}0.5).
This reinforces our earlier observations regarding gender differences, highlighting the critical role of phone usage in male students' mental health, as it appears to significantly increase the risk of severe stress and anxiety.
These preliminary findings on the gender differences suggest that the mechanisms linking phone usage to mental health may vary significantly across groups. Consequently, predictive models for mental health based on phone behaviors should be tailored for specific groups—or even individuals—to capture these variations effectively.

\textbf{The Effects of Location Contexts}. Table~\ref{logit} further shows the regression results in different locations, 
including food places, social places, study places, home, and dormitories,  revealing significant contextual differences. For instance, unlock duration is positively associated with the likelihood of falling into the “Severe” group at social places, study places, food places, and home but shows a negative association in dormitories. 
These variations underscore the importance of incorporating context as a critical dimension in predictive models. Future models should be context-aware, accounting for location-specific behaviors to improve accuracy and relevance in predicting mental health outcomes.

\vspace{-2pt}
\section{Related Work}\label{literature}
\vspace{-2pt}
This paper is the first study to investigate college students' smartphone unlocking behaviors and mental well-being using a longitudinal dataset spanning entire college years, including both iOS and Android users.
Previous research has explored unlocking behaviors through indirect methods, such as examining eye gaze during unlocking~\cite{abdrabou2024eyegaze}, conducting self-reported online surveys~\cite{harbach2014survey}, predicting unlock patterns based on swipe behaviors~\cite{li2019swipevlock}, and analyzing lock screens on Android devices only~\cite{harbach2016anatomy}. However, none of these studies have directly measured real-world unlocking behaviors in practice.

Using mobile sensing techniques to understand mental health issues has gained prominence, such as identifying mental health risks~\cite{adler2022machine, faurholt2019objective, wang2022first}, modeling behaviors and capturing mental health parameters~\cite{likamwa2013moodscope, macleod2021mobile, mehrotra2017mytraces, tag2022emotion}. While some studies focus on college students' behaviors~\cite{macleod2021mobile, meegahapola2020smartphone, meegahapola2023generalization, wang2021transition}, the specific role of smartphone unlocking behaviors remains underexplored.


 Another area has linked smartphone usage to diminishing well-being, with studies examining their relationship~\cite{buchi2024digital, allcott2022digital, peper2018digital, thomas2022digital} and interventions to reduce digital usage~\cite{lyngs2020just, orzikulova2023finerme, roffarello2023achieving, zimmermann2023digital}. 
 However, they often overlook that smartphone behaviors involve both \textit{duration} and \textit{frequency}, with patterns varying across different user groups and contexts. Our work fills this gap by providing insights to inform the design of personalized, context-aware interventions for healthier smartphone usage.


\vspace{-2pt}
\section{Conclusion and Future Directions}\label{conclusion}
\vspace{-2pt}
This study presents the first large-scale investigation of college students' smartphone unlocking behaviors and their association with mental well-being, leveraging over 210,000 data points from a four-year longitudinal dataset. Our findings highlight significant gender and contextual differences, emphasizing the need for fine-grained analyses in mental health research. We demonstrate that unlocking behaviors alone offer predictive power, enabling lightweight models for practical adoption.
Future work should refine predictive models beyond gender differences, incorporate temporal dynamics such as high-stress periods, and develop personalized, context-aware interventions, including location-based recommendations, to promote healthier phone usage and digital well-being.

\bibliographystyle{IEEEtran}
\bibliography{ref}

\end{document}